\documentclass[aps,amsmath,amssymb,reprint,superscriptaddress]{revtex4-2}

\usepackage{graphicx}           
\usepackage{dcolumn}            
\usepackage{bm}                 
\usepackage{hyperref}           
\usepackage[mathlines]{lineno}  
\usepackage{upgreek}            
\usepackage{pifont}             
\usepackage[dvipsnames]{xcolor} 
\usepackage{booktabs}
\usepackage{tabularx}
\usepackage{array}

\newcolumntype{A}{>{\hspace*{-2.5em}\centering\arraybackslash}X}
\newcolumntype{B}{>{\hspace*{-2.8em}\centering\arraybackslash$}X<{$}}
\newcolumntype{C}{>{\hspace*{-2.1em}\centering\arraybackslash$}X<{$}}
\newcolumntype{D}{>{\hspace*{-1.8em}\centering\arraybackslash$}X<{$}}
 
\begin{document}

\preprint{APS/123-QED}

\title{Suppressing Cavity Frequency Noise Using a Kerr Nonlinearity}

\author{J.P.~van Soest}
\email{j.vansoest-1@tudelft.nl}
\affiliation{Kavli Institute of NanoScience, Delft University of Technology, PO Box 5046, 2600 GA Delft, Netherlands}

\author{S.~Meilof}
\affiliation{Kavli Institute of NanoScience, Delft University of Technology, PO Box 5046, 2600 GA Delft, Netherlands}

\author{G.L.~Bhai}
\affiliation{Kavli Institute of NanoScience, Delft University of Technology, PO Box 5046, 2600 GA Delft, Netherlands}

\author{M.~Villiers}
\affiliation{Kavli Institute of NanoScience, Delft University of Technology, PO Box 5046, 2600 GA Delft, Netherlands}

\author{C.A.~Potts}
\affiliation{Department of Electrical and Software Engineering, University of Calgary, 2500 University Drive NW, Calgary, Alberta T2N 1N4, Canada}

\author{G.A.~Steele}
\email{g.a.steele@tudelft.nl}
\affiliation{Kavli Institute of NanoScience, Delft University of Technology, PO Box 5046, 2600 GA Delft, Netherlands}
 
\begin{abstract}
Resonance-frequency fluctuations can limit the sensitivity and stability of superconducting microwave cavities used for qubit readout, optomechanical displacement sensing, and magnetic flux detection. Here, we demonstrate the suppression of resonance-frequency fluctuations by locking a noisy nonlinear superconducting microwave cavity to a strong pump tone. The Kerr nonlinearity of this system, whereby the resonance frequency depends on the intracavity field amplitude, gives rise to an intrinsic feedback mechanism that enables passive stabilization without active external feedback. Using two-tone spectroscopy, we experimentally characterize the intrinsic nonlinear feedback mechanism and investigate its temporal stability through Allan deviation analysis. The frequency fluctuations of the locked cavity mode are reduced by nearly two orders of magnitude, reaching the $1/f$~noise floor, without continuous frequency tracking or active control. Kerr locking provides a general approach for self-stabilizing nonlinear microwave resonators by suppressing low-frequency cavity noise while preserving sensitivity to signals outside the locking bandwidth. This approach may benefit a broad range of systems, including SQUID-based resonators, optomechanical devices, and parametric amplifiers.
\end{abstract}

\maketitle

\section{Introduction} \label{Intro}
Superconducting microwave cavities are widely employed as resonant sensors for applications including qubit readout, optomechanical displacement sensing, and magnetic flux detection \cite{Krantz2019, blais2021circuit, aspelmeyer2014cavity, hatridge2011dispersive}. In these applications, the measured quantity typically manifests as a shift of the cavity resonance frequency, which is detected through the reflected microwave field \cite{regal2008measuring, Reed2010, wallraff2005approaching}. However, environmental perturbations, including mechanical vibrations and magnetic flux noise, can produce additional low-frequency fluctuations of the cavity resonance \cite{beysengulov2022helium, Bothner2022, Potts2025}. When these fluctuations become comparable to or exceed the cavity linewidth, conventional fixed-frequency readout loses sensitivity and becomes unreliable \cite{brock2020frequency}.

A method that suppresses slow fluctuations in resonance frequency while retaining sensitivity to faster signals is therefore desirable. Active resonance-tracking techniques, such as Pound–Drever–Hall locking and related real-time feedback schemes, can compensate for slow frequency fluctuations by estimating the instantaneous cavity detuning and adjusting the microwave probe accordingly \cite{vanSoest2023, lindstrom2011pound, kanhirathingal2022feedback}. However, these methods require continuous measurement of the cavity response and an external feedback path to control the probe frequency. An alternative is to exploit the cavity's intrinsic nonlinear dynamics to generate the required feedback. In particular, many superconducting microwave cavities exhibit a Kerr nonlinearity that makes their resonance frequency dependent on the intracavity field amplitude \cite{Siddiqi2005, tholen2007nonlinearities, ong2011circuit}. 

In this article, we demonstrate that this nonlinearity can be used to passively stabilize the cavity response without measurement-based feedback. A strong, fixed-frequency pump generates a pair of symmetric pump-dressed Bogoliubov modes \cite{Sani2021}. Changes in the undressed cavity resonance modify the intracavity field and, hence, the Kerr frequency shift. This intrinsic nonlinear response thereby suppresses fluctuations in the dressed-mode frequencies. We demonstrate this mechanism experimentally in a superconducting cavity subject to frequency noise substantially larger than its linewidth. Moreover, we investigate the dynamics of the locking process and characterize its long-term stability using the Allan deviation. The frequency fluctuations of the stabilized mode are reduced by nearly two orders of magnitude, reaching the $1/f$~noise floor without continuous frequency tracking or external feedback.
\begin{figure}[t!]
    \centering
    \includegraphics[width=\linewidth]{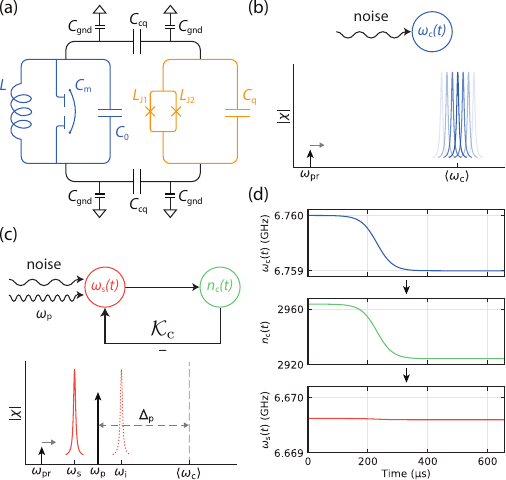}
    \caption{\textbf{Illustration of the Kerr-feedback mechanism in a microwave resonator.} (a) Schematic representation of the circuit. The cavity is shown in blue, the qubit in orange. (b) Diagram of a cavity resonance subject to environmental noise that fluctuates the resonance frequency $\omega_{\rm c}\!\left( t \right)$ (blue). A small probe tone $\omega_{\rm pr}$ (black arrow) is swept over the fluctuating cavity resonance. (c)  A pump tone with frequency $\omega_{\rm p}$ (large black arrow) is applied red-detuned from the bare cavity frequency $\left< \omega_{\rm c} \right>$. The linearized response of the Kerr cavity is described by the dressed cavity resonance $\omega_{\rm s}\!\left( t \right)$ (solid red), and an additional idler resonance $\omega_{\rm i}\!\left( t \right)$ (dotted red), located symmetrically about the pump frequency. Fluctuations of the bare cavity resonance result in a time-dependent intracavity photon number $n_{\rm c}\!\left( t \right)$ (green). The dressed cavity resonance is stabilized by an intrinsic negative feedback mechanism arising from the cavity Kerr nonlinearity $\mathcal{K}_{\rm c}$. (d) Numerical simulation of the temporal response of the intracavity photon number to a smooth step in the underlying cavity resonance frequency. The induced change in photon number compensates for the shift of the bare cavity, stabilizing the locked dressed cavity frequency.}
    \label{Figure1}
\end{figure}%
\section{Device and Operating Principle} \label{Setup}
Our device consists of a superconducting microwave resonance circuit on two chips in a flip-chip geometry, a schematic of which is shown in Fig.~\ref{Figure1}(a). The top chip contains a high-tensile-stress silicon nitride membrane embedded within an in-substrate phononic shield \cite{vanSoest2023}. The membrane is metallized with $250 \, {\rm \upmu m} \times  250 \, {\rm \upmu m} \times 50 \, {\rm nm}$ evaporated aluminum. At the core of the device is a lumped-element microwave cavity coupled to a transmission line. It consists of a meander inductor ($L$) and two capacitor plates ($C_{\rm 0}$) etched into a $ 100$~nm-thick NbTiN film on a silicon substrate. The cavity, in the absence of the top chip, is designed to have a resonant frequency of $\omega_{\rm bot} / 2\pi = 19.65$~GHz. As the top chip is brought in proximity to the bottom chip, the membrane loads the circuit capacitance ($C_{\rm m}$) and reduces the frequency to $\omega_{\rm c} / 2\pi = 6.76$~GHz. To selectively introduce nonlinearity, a flux-tunable floating double-island transmon qubit is coupled to the lumped-element microwave cavity \cite{Vool2017}. The qubit frequency can be controlled with an external magnetic coil, which threads a magnetic flux through the $125 \ {\rm \upmu m^{2}}$ loop of the asymmetric superconducting quantum interference device (SQUID). For the simulated dipole coupling strength of $g_{\rm cq} / 2\pi = 1$~MHz, the qubit induces a self-Kerr nonlinearity in the cavity of 
\begin{equation} \label{Eq_cav_nonl}
    \mathcal{K}_{\rm c} = \mathcal{K}_{\rm q} \cdot \left( \frac{g_{\rm cq}}{\Delta_{\rm cq}} \right)^{4},
\end{equation}
where $\mathcal{K}_{\rm q} = - E_{\rm c}/\hbar$ is the self-Kerr nonlinearity of the qubit, dependent on its charging energy $E_{\rm c}$ \cite{blais2021circuit}, and $\Delta_{\rm cq}$ is the qubit's detuning from the cavity. The qubit is uncharacterized in this work; nevertheless, the induced cavity Kerr coefficient is determined to be $\mathcal{K}_{\rm c} / 2\pi = -24.1$~kHz; see Supplementary Material Sec.~F. The flip-chip device is clamped with spring-loaded screws and mounted vertically on a mass-spring system that hangs below the base plate of a dilution refrigerator ($T \approx 20$~mK). Further details of the device fabrication and the measurement setup are provided in Supplementary Material Secs.~A~and~B.

As described above, in many practical implementations, superconducting microwave cavities exhibit resonance-frequency fluctuations that limit their performance \cite{Gao2007, burnett2014evidence}. Figure~\ref{Figure1}(b) illustrates how noise can result in large fluctuations of the bare cavity frequency $\omega_{\rm c}\!\left( t \right)$. These fluctuations can significantly increase the time-averaged cavity linewidth $\kappa$. Under a strong drive from a second tone, the linearized response of the Kerr cavity is described by two coupled Bogoliubov modes, which appear as a dressed cavity resonance, sometimes referred to as the signal mode, along with a second mirrored image mode, sometimes referred to as an idler mode \cite{Sani2021} or Bogoliubov ghost  \cite{Frerot2023}, symmetric about the pump frequency, as shown in Figure~\ref{Figure1}(c). The detuning of these quasi-modes relative to the pump is determined by the intracavity occupation of the pump tone and results in an intrinsic proportional negative feedback when the pump is blue-detuned from the dressed cavity mode, the details of which are discussed below. A shift in the bare resonance frequency modifies the intracavity occupation together with the pump detuning; this photon-number change acts as a feedback loop, stabilizing the frequency of the dressed cavity resonance $\omega_{\rm s}\!\left( t \right)$, shown in Fig.~\ref{Figure1}(d).

\section{Kerr Locking the Nonlinear Cavity} \label{Locking}

The origin of the intrinsic feedback can be understood from the linearized response of a pumped Kerr resonator. For a pump detuning ($\Delta_{\rm p}=\omega_{\rm p}-\langle\omega_{\rm c}\rangle$), the steady-state intracavity occupation is determined by
\begin{equation}
    n_{\rm c}\bigg[ (\Delta_{\rm p} - \mathcal{K}_{\rm c}n_{\rm c})^2 + (\kappa/2)^2  \bigg] = (\kappa_{\rm ext}/2)\vert a_{\rm in} \vert^2,
    \label{Eq_PhotonNum}
\end{equation}
where $n_{\rm c}$ is the intracavity photon number, $\kappa$ is the total cavity linewidth, $\kappa_{\rm ext}$ is the cavity coupling rate, and $a_{\rm in}=\sqrt{\mathcal{P}/\hbar\omega_{\rm p}}$ is the incident photon flux for a power $\mathcal{P}$ launched into the cavity.

Linearizing the cavity dynamics about this driven state produces two modes, which appear in the response to a weak probe as symmetric pump-dressed Bogoliubov resonances \cite{Sani2021, Yamaji2022}. Their frequencies are given by the real-valued function
\begin{equation} \label{Eq2}
     \omega_{\rm \pm} = \omega_{\rm p} \pm \sqrt{(\Delta_{\rm p} - \mathcal{K}_{\rm c}n_{\rm c})(\Delta_{\rm p} - 3\mathcal{K}_{\rm c}n_{\rm c})}.
\end{equation}
In the presence of a strong drive, both modes are always present in the response to a weak probe. However, the spectroscopically bright signal mode, with frequency $\omega_{\rm s}$, is continuously connected to the bare cavity resonance; it will therefore be our primary focus \cite{Bothner2022,Sani2021}.

Equations~\ref{Eq_PhotonNum} and \ref{Eq2} describe the intrinsic feedback mechanism. A fluctuation of the bare cavity frequency changes the instantaneous pump detuning. This changes the intracavity occupation and, through the Kerr nonlinearity, the frequency of the dressed cavity resonance. Depending on the operating point, this pump-induced shift either reinforces or opposes the original cavity-frequency fluctuation. This gives rise to three relevant regimes.

First, for a blue-detuned pump, ($\Delta_{\rm p}>0$), Eq.~\ref{Eq_PhotonNum} has a single stable solution. In this regime, a fluctuation of the bare cavity frequency changes $n_{\rm c}$ such that the resulting Kerr shift opposes the original fluctuation. The dressed cavity resonance therefore fluctuates less than the bare cavity, producing intrinsic frequency stabilization. However, in this pumping regime the intracavity photon number remains small, and therefore the feedback is weak. 

\begin{figure}[t!]
    \centering
    \includegraphics{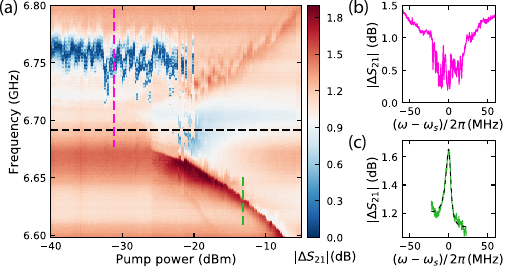}
    \caption{\textbf{Observation of the Kerr-locked cavity.} (a) Normalized colormap of the magnitude $|\Delta S_{21}|$ response, measured using two-tone spectroscopy. A pump tone, red-detuned from the bare cavity by $\Delta_{\rm p} / 2\pi = -68$~MHz (black dashed line), is progressively increased in power while a weak, non-invasive probe tone is swept across the cavity resonance. At low pump powers, fluctuations of the cavity resonance are observed about $\left< \omega_{\rm c} \right> / 2\pi = 6.76$~GHz. Above a pump power of $-20.0$~dBm the nonlinear response bifurcates, resulting in the dressed cavity resonance to appear below the pump frequency. This mode is consequently frequency-stabilized, together with the idler mode on the opposite side of the pump. The suppression of resonance-frequency fluctuations is attributed to the intrinsic negative feedback provided by the cavity Kerr nonlinearity $\mathcal{K}_{\rm c}$. Representative spectra acquired before (b) and after (c) the onset of the Kerr locking, corresponding to the similarly colored dashed lines in panel (a). The frequency axes are centered around the (average) dressed cavity frequency $\omega_{\rm s}$ for each resonance. A Lorentzian fit (black dashed line) to the locked dressed cavity resonance provides a linewidth of $\kappa_{\rm s}/ 2\pi = 7.6$~MHz, unperturbed by the bare cavity fluctuations.}
    \label{Figure2}
\end{figure}

Second, for a red-detuned pump ($\Delta_{\rm p}<0$), below the bifurcation, the cavity occupies the stable low-photon-number branch. The pump-induced AC Stark shift pulls the dressed cavity resonance towards the pump, but the corresponding response reinforces fluctuations of the bare cavity frequency. The cavity therefore remains unlocked, and its frequency noise is amplified rather than suppressed. In our experiment, the pump is applied at $\Delta_{\rm p}/2\pi=-68~\mathrm{MHz}$. At low pump powers, the resulting frequency excursions substantially exceed the intrinsic cavity linewidth, producing the broadened response shown in Fig.~\ref{Figure2}(a,b).

Third, as the red-detuned pump power is increased, Eq.~\ref{Eq_PhotonNum} becomes multistable. The system crosses a saddle-node bifurcation and enters the stable high-photon-number branch \cite{andersen2020quantum,Siddiqi2005}. Near the exceptional point, $n_{\rm c}=\Delta_{\rm p}/3\mathcal{K}_{\rm c}$, the feedback changes sign. The pump is then blue-detuned from the dressed cavity resonance, even though it remains red-detuned from the bare cavity, see Fig.~\ref{Figure2}(a). Consequently, the pump-induced Kerr shift opposes fluctuations of the bare cavity frequency and the dressed cavity resonance becomes locked. Thus, the blue-pumped regime and the high-photon-number branch of the red-pumped regime produce locking through the same underlying condition: the pump lies blue-detuned relative to the dressed cavity resonance at frequency $\omega_-$.

A representative trace of the locked dressed cavity resonance is shown in Fig.~\ref{Figure2}(c). A Lorentzian fit gives ($\kappa_{\rm s}/2\pi=7.6~\mathrm{MHz}$), allowing the underlying linewidth to be resolved despite substantially larger low-frequency fluctuations of the bare cavity. Both the signal and idler resonances appear as amplification peaks. Reflection above unity indicates that constructive interference from the four-wave-mixing process provides sufficient parametric amplification to overcome the internal cavity losses \cite{Huber2020,Sani2021}. We observe a maximum output gain of approximately $1~\mathrm{dB}$.
\section{Time Dynamics of the Kerr-Feedback} \label{Time}
\begin{figure}[t!]
    \centering
    \includegraphics{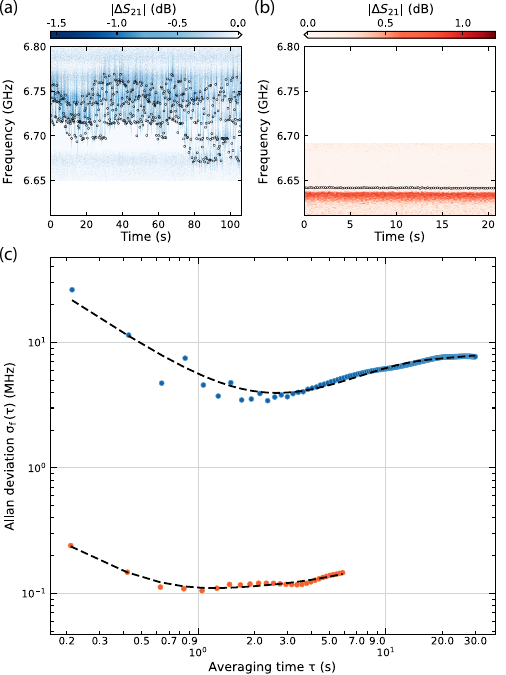}
    \caption{\textbf{Time dynamics of the cavity resonance before and after Kerr-feedback.} (a) Normalized colormap of the measured $|\Delta S_{21}|$ response of fast spectroscopy traces. The unlocked bare cavity displays resonance-frequency fluctuations and linewidth broadening. The fit resonance frequency of each trace is shown with black circles, exhibiting a standard deviation of $\sigma_{\rm std,c} = 23.2$~MHz. Measurement and fits of the dressed cavity resonance are shown in panel (b), where the standard deviation is reduced to $\sigma_{\rm std,s} = 297$~kHz. (c) Absolute frequency Allan deviation $\sigma_{\rm f}\!\left( \tau \right)$ before (blue) and after (red) applying the intrinsic feedback mechanism to the cavity. Black dashed lines show fits based on the noise model. The total frequency noise of the cavity is reduced by nearly two orders of magnitude for the locked dressed cavity resonance.}
    \label{Figure3}
\end{figure}%
To quantify the mitigation of cavity-frequency noise, the time dynamics of the feedback mechanism are investigated. Fast single-tone spectroscopy traces of the unlocked cavity were measured, as shown in Figure~\ref{Figure3}(a). Individual traces were recorded at $213$~ms intervals, which is shorter than the timescale of the dominant fluctuations, yielding snapshots of the cavity resonance. The measurement was repeated for the locked dressed cavity resonance, shown in Figure~\ref{Figure3}(b). Each trace was fit with a Lorentzian, and the extracted resonance frequencies are shown as black circles. As the bare cavity frequency fluctuations are mitigated by the Kerr nonlinearity, the dressed cavity frequency is stabilized in time. The standard deviation of the unlocked cavity is $\sigma_{\rm std,c} = 23.2$~MHz. For the locked cavity, this is reduced to $\sigma_{\rm std,s} = 297$~kHz, which is much smaller than the cavity linewidth, $\sigma_{\rm std,s} \ll \kappa_{\rm s}$.

The nature of the cavity noise is investigated by computing the overlapping Allan deviation of the fractional resonance-frequency fluctuations, $\delta \omega_{\rm c} / \langle \omega_{\rm c} \rangle$, extracted from the fits described above. Figure~\ref{Figure3}(c) shows the Allan deviation $\sigma_{\rm f}\!\left( \tau \right)$ converted to absolute frequency units, for the unlocked bare cavity (blue) and locked dressed cavity resonance (red). The black dashed lines indicate fits based on the noise model discussed in Supplementary Material Sec.~H. For short averaging times $\tau$, the measured fluctuations are dominated by phase noise changing the instantaneous cavity frequency, as evidenced by the $\tau^{-1}$ scaling in the log-log representation of the Allan deviation. As $\tau$ increases, the Allan deviation starts to rise with a slope~$\frac{1}{2}$ due to slow Brownian frequency fluctuations of the cavity, as commonly observed in resonator systems \cite{harada2013slow}. The unlocked bare cavity does not reach the pink-noise limit because this contribution is overwhelmed by other noise processes. However, the Allan deviation of the locked dressed cavity resonance is substantially smaller. With the Kerr-feedback, $\sigma_{\rm f}$ reaches a plateau at intermediate $\tau$, consistent with a $1/f$~noise floor that is likely dominated by instrumental noise. Overall, the total cavity frequency noise is reduced by nearly two orders of magnitude.

\section{Discussion \& Outlook} \label{Conclusion}
We have demonstrated the suppression of cavity resonance-frequency fluctuations by Kerr locking a noisy microwave cavity to a strong pump tone. A floating transmon qubit was coupled to a flip-chip microwave cavity, thereby inducing a self-Kerr nonlinearity in the cavity mode. Magnetic and vibrational fluctuations cause the cavity resonance frequency, and consequently the intracavity photon number, to vary in time. When a strong pump tone is applied blue-detuned from the dressed cavity frequency, the Kerr cavity exhibits a linearized response described by two coupled Bogoliubov modes. These resonances appear symmetrically positioned about the pump frequency, with a separation that depends on the intracavity occupation of the pump tone. Changes in the undressed cavity resonance modify the intracavity field and, hence, the Kerr frequency shift, resulting in an intrinsic negative-feedback mechanism. This stabilization allows the unperturbed cavity linewidth to be observed.

By analyzing the time dynamics of this feedback mechanism, we evaluated the Allan deviation for both the unlocked bare cavity and the Kerr-locked dressed cavity resonance. The unlocked cavity is dominated by phase and Brownian noise contributions that substantially exceed the flicker-noise floor, whereas the fluctuations in the dressed-mode frequencies are suppressed by nearly two orders of magnitude, reaching the instrumental $1/f$~noise floor.

The Kerr locking mechanism is governed by a set of highly nonlinear equations. In Supplementary Material Sec.~G, numerical simulations based on this model reproduce the observed stabilization of the nonlinear cavity and show that the feedback bandwidth is on the order of the cavity linewidth, ${\sim}\kappa$. Remarkably, modulations of the cavity resonance frequency at rates exceeding the cavity linewidth are predicted to be amplified by the Kerr locking process, suggesting a potential route toward enhanced detection of mechanical signals in optomechanical devices. Future theoretical work could include developing an effective linear feedback model, which is beyond the scope of this work. 

More broadly, Kerr locking is applicable to stabilizing nonlinear microwave resonators by suppressing low-frequency cavity noise while preserving sensitivity to signals outside the locking bandwidth. These capabilities may be relevant to a wide range of platforms, including SQUID-based resonators and parametric amplifiers. Recent studies have demonstrated Kerr-enhanced backaction \cite{Zoepfl2023, Diaz-Naufal2025} and enhancing flux-mediated optomechanical interactions \cite{Schmidt2020, Rodrigues2021, Bothner2022, Rodrigues2022, Luschmann2022, Schmidt2024}. Moreover, Kerr optomechanics could be used for intracavity squeezing \cite{Yurke2006, Boutin2017, Monsel2021} and quantum transduction \cite{Lauk2020, Mirhosseini2020}, indicating that Kerr locking may provide a powerful approach for controlling and exploiting nonlinear dynamics in these systems. In other systems involving these quasi-modes, level attraction was observed, creating exceptional points \cite{Eleuch2014, Bernier2014, Bernier2018, Miri2019, Sani2021}. Finally, coupling an optomechanical cavity to a qubit has been proposed as a platform for preparing mechanical quantum states \cite{Khosla2018, Bergholm2019, Kounalakis2019, Kounalakis2020, Hauer2023, Potts2025, Gerashchenko2025}.
 
\subsection*{Author Contributions}
J.P.S.\ contributed to the conceptual development, device design and simulation, fabrication, fridge operation, measurements, data analysis, theory and numerical simulation, and writing of the manuscript.
S.M.\ contributed to the analytical theory and numerical simulations, data analysis and fitting, the fabrication recipe, and gave feedback on the figures and text of the manuscript. 
G.L.B.\ contributed to the experimental work, including measurements, measurement planning, fridge operation, and data acquisition, contributed to the conceptual development, and reviewed the manuscript.
M.V.\ provided support with microwave simulations.
C.A.P.\ co-supervised the project and contributed to conceptual development, numerical simulations, theoretical analysis, and manuscript writing.
G.A.S.\ supervised the project, proposed the idea of implementing Kerr stabilization, helped develop the conceptual understanding and formulate the storyline, and gave feedback on the figures and text of the manuscript.

\begin{acknowledgments}
We thank M. Arfini, R.C. Dekker, and S. Deve for their assistance with device fabrication. This publication is part of the project ‘Superconducting Electromechanics: Massive superpositions for exploring quantum mechanics and general relativity’, project number VI.C.212.087 of the research program VICI round 2021, financed by the Dutch Research Council (NWO). S.M. acknowledges the Kavli Foundation under the Kavli Institute Innovation Award (KIIA) (LS-2023-GR-14-2778) and the Kavli Institute of Nanoscience Delft. C.A.P. acknowledges the support of the Natural Sciences and Engineering Research Council of Canada (NSERC) (No.~RGPIN-2026-05057).
\end{acknowledgments}


\bibliography{References.bib}

\clearpage

\onecolumngrid

\begin{center}
    {\LARGE\bfseries Supplementary Material\par}
    \vspace{0.5em}
    {\large Suppressing Cavity Frequency Noise Using a Kerr Nonlinearity\par}
\end{center}

\vspace{1em}

\twocolumngrid

\maketitle
\tableofcontents

\section{Device Fabrication} \label{SI_Fab}
The device used in the work was fabricated in the cleanroom of the \textit{Kavli Nanolab} at \textit{Delft University of Technology} \cite{Kavli_Nanolab}. The fabrication process consists of two parts: the bottom microwave chip and the top membrane chip. These two parts are combined in our flip-chip holder to complete the device. In the following section, we describe the fabrication and assembly steps in detail.

\subsection{Microwave Chip}
\begin{figure*}[t!]
    \centering
    \includegraphics[width=0.8675\textwidth]{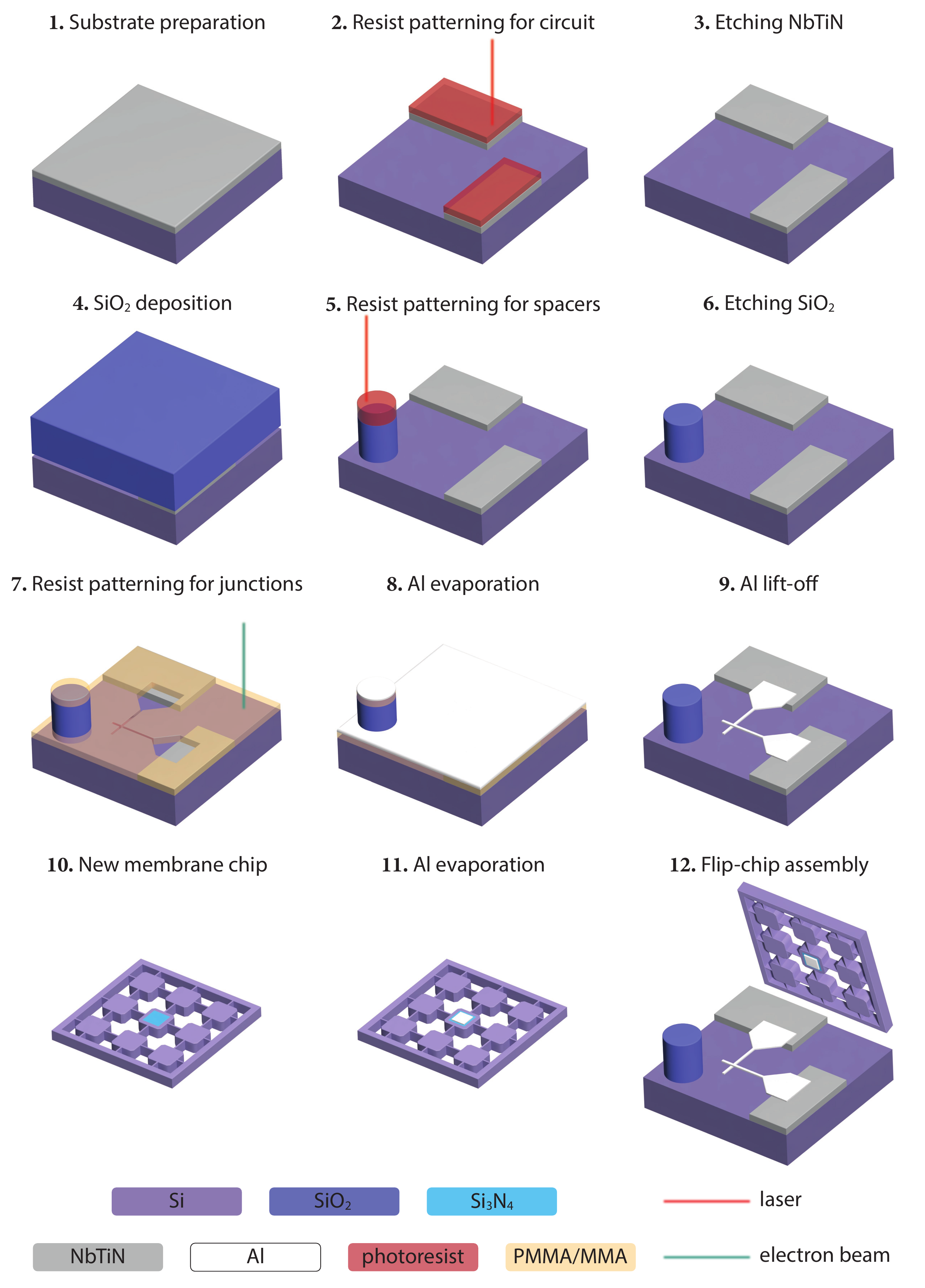}
    \caption{\textbf{Schematic device fabrication.} Panels \textbf{1-3} show the fabrication of the NbTiN microwave circuit. Panels \textbf{4-6} show the deposition and etching of the ${\rm SiO_{2}}$ spacers. Panels \textbf{7-9} show the evaporation and lift-off process for the Al junctions. Panels \textbf{10-12} show the metallization and placement of the membrane chip. Dimensions of the structures are not to scale. Detailed information on each step is provided in the text.}
    \label{Figure_SI_recipe}
\end{figure*}

A schematic overview of the recipe for the microwave chip is shown in Fig.~\ref{Figure_SI_recipe}.\\ 
\\
\textbf{Step 1: Substrate preparation.} We start the process with a $500 \, {\rm \upmu m}$ thick high-resistivity Si wafer, on which a $100$~nm thick NbTiN film has been deposited \cite{Thoen2017}, supplied by the \textit{Netherlands Institute for Space Research (SRON)} \cite{SRON}. The wafer is diced into $13 \times 13$~mm chips, after which they are cleaned with PRS3000 and rinsed with acetone and IPA.\\
\\
\textbf{Step 2: Resist patterning for the circuit.} For the patterning of the microwave circuit and two sets of alignment markers, we use optical lithography. As we will have to align the $10 \times 10$~mm membrane chip with respect to the $13 \times 13$~mm microwave chip, this first alignment step needs to be precise in angle and position. We create a mask by spinning a $400$~nm thick layer of S1805 photoresist. The resist mask is exposed with a $365$~nm laser beam with a dose of $120 \, {\rm mJ/cm^{2}}$. Afterwards, we develop the sample in the developer MF-21A for $70$~seconds, after which we proceed with sequentially $40$~seconds and $120$~seconds in two beakers of ${\rm H_{2}O}$.\\
\\
\textbf{Step 3: Reactive ion etching $\mathbf{NbTiN}$.} To create the circuit structures in the metal film, we perform a ${\rm CF_{4}}$ reactive ion etch (RIE). At an etch rate of $1.4 \, {\rm nm/sec}$. The process is applied for a total of $78$~seconds, of which $6$~seconds is over-etch. Laser end-point detection is used to prevent over-etching the substrate, as this affects the adhesion of the aluminum junctions, which are discussed in Step 8. To prevent the creation of fluorocarbons, an \textit{in situ} oxygen plasma is performed after the metal etch, before breaking the vacuum. The leftover resist is stripped with the solvent PRS3000 and cleaned by the same process described above. \\
\\
\textbf{Step 4: $\mathbf{SiO_{2}}$ deposition.} We fabricate spacers to prevent the membrane chip from crashing into the bottom chip during the assembly of the flip-chip device. This also aids the parallelism of the two chips. We deposit a layer of ${\rm SiO_{2}}$ over the entire chip using plasma-enhanced chemical vapor deposition (PECVD). ${\rm SiO_{2}}$ is deposited at a rate of $1.2 \, {\rm nm/sec}$ and the dielectric layer grows $900$~nm tall in 12:52~minutes. After the deposition, we clean the chip again with acetone and IPA.\\
\\
\textbf{Step 5: Resist patterning for the spacers.} For the second patterning step, we spin a $500$~nm thick layer of S1805 photoresist on the chip. Next, we align to the first set of markers in the laser writer, after which the resist is exposed in two steps. First, the spacers are written with a dose of $120 \, {\rm mJ/cm^{2}}$. Second, the edge of the chip is written again with a dose of $160 \, {\rm mJ/cm^{2}}$. This is done in order to counter any edge effects, which could result in leftover ${\rm SiO_{2}}$ close to the launchpads and feedlines of the microwave circuit. The exposed resist is developed using MF-21A following the same recipe as before.\\
\\
\textbf{Step 6: Buffered oxide etching $\mathbf{SiO_{2}}$.} The ${\rm SiO_{2}}$ layer is removed with a buffered oxide etch (BOE). The solution consists of 7:1 volume ratio of $40\%$ ammonium fluoride $\left( {\rm NH_{4}F} \right)$ and $49\%$ hydrofluoric acid $\left( {\rm HF} \right)$. The latter etches the ${\rm SiO_{2}}$ layer. We use the BOE-solution rather than pure ${\rm HF}$ to retain a constant ${\rm HF}$ concentration and thereby a constant etch rate \cite{Burham2016}. The etch rate for the deposited ${\rm SiO_{2}}$ is $1.1 \, {\rm nm/sec}$, which removes the $900$~nm layer in 13:30~minutes. We rinse the chip twice with ${\rm H_{2}O}$ before the leftover resist is stripped in PRS3000 with the same recipe as before. ${\rm SiO_{2}}$ being a lossy dielectric material can reduce the quality factor of the microwave resonators. The spacers are therefore designed to be far removed from the microwave circuit.\\
\\
\textbf{Step 7: Resist patterning for the junctions.} The last patterning step of the microwave chip is to create the Josephson junctions of the superconducting quantum interference device (SQUID) of the transmon qubit. To obtain the resolution needed for junctions with an overlap area on the order of $100$~nm we use electron beam lithography (EBL). First, we prepare the substrate by performing an oxygen plasma to clean the surface. Then, we build a bi-layer resist stack by first spinning a $180$~nm thick layer of MMA $8.5\%$ EL6, followed by a $500$~nm thick layer of PMMA 950K A6. In the lithography setup, the chip is aligned to the second set of precise markers. We pattern the junctions using an e-beam spot size of $5$~nm and a dose of $1700 \, {\rm \upmu C/cm^{2}}$. The resist stack is then developed in a cold $\left( 7 \, {\rm ^{\circ}C} \right)$ ${\rm IPA:H_{2}O}$ 3:1 solution in a sonicator at the lowest setting for $2$~minutes. Afterwards, the chip is transferred to an IPA bath for $15$~seconds.\\
\\
\begin{figure}[t!]
    \centering
    \includegraphics[width=0.4\textwidth]{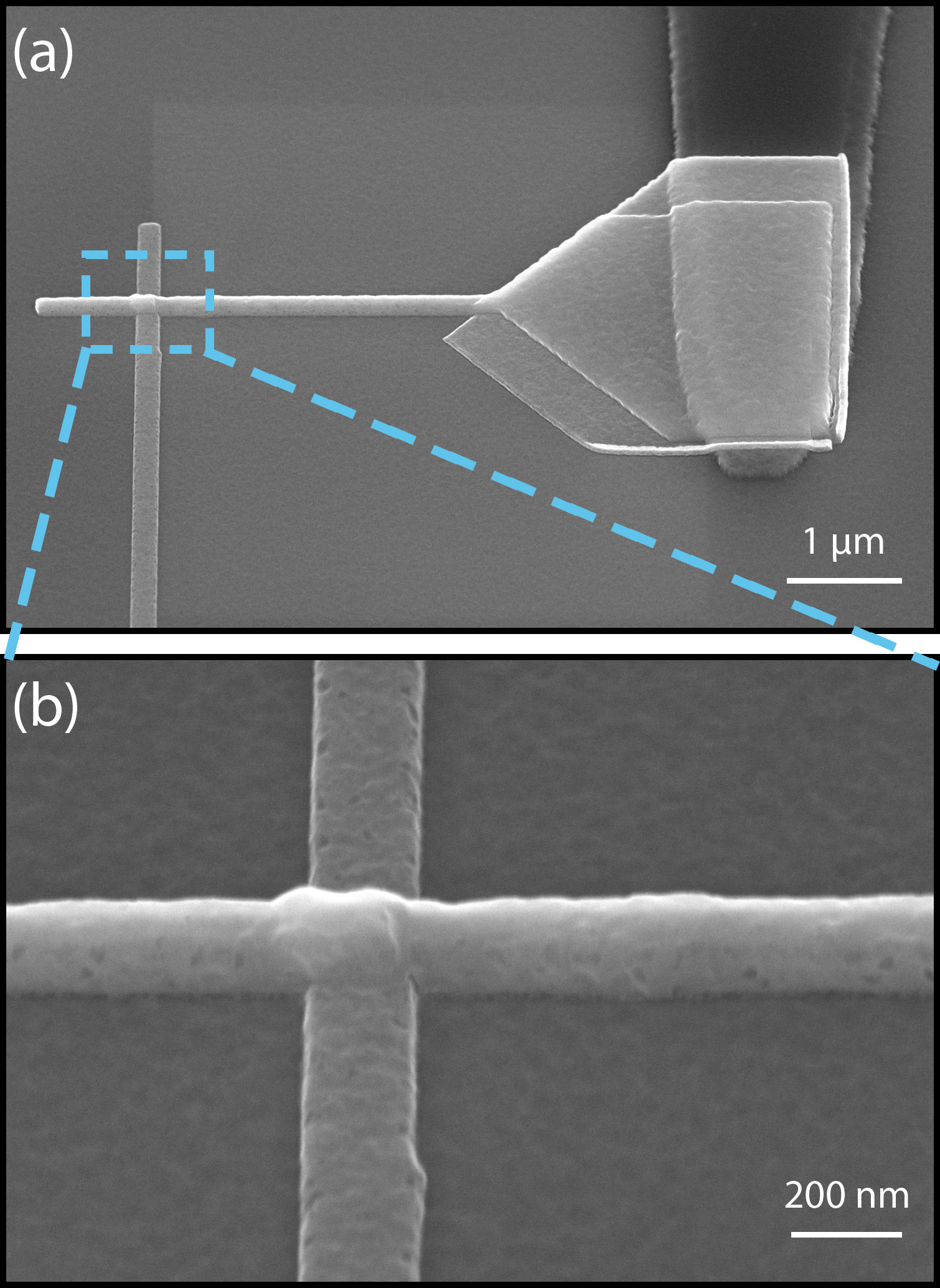}
    \caption{\textbf{Scanning electron micrographs of an aluminum junction.} Two different magnifications of a Josephson junction fabricated with the described recipe. The images are taken with a tilt of $40^{\circ}$. Panel (a) shows the intact step edges of the Al (white) from the Si substrate (gray) to the NbTiN circuit (black). (b) A zoom-in of the Manhattan-style Al-AlOx-Al Josephson junction.}
    \label{Figure_SI_junction}
\end{figure}%
\textbf{Step 8: Aluminum evaporation.}
In order to ensure a clean surface, we prepare the sample before the Al evaporation. First, we again perform an oxygen plasma. Then we do a short submersion in a BOE-solution: ${\rm H_{2}O}$ for $10$~seconds, 7:1 BOE for $1$~minute, twice ${\rm H_{2}O}$ for $1$~minute, ${\rm H_{2}O}$ for $5$~minutes. After the acid clean, the chip is immediately mounted on the evaporator cassette and placed in the vacuum loadlock of the evaporator. By applying titanium sublimation, the pressure in the main vacuum chamber is brought to $1.1 \cdot 10^{-7}$~mbar, where we can start the deposition of the junctions. The Manhattan-style junctions are deposited using double-angle shadow evaporation \cite{Costache2012}. The first layer of $35$~nm thick Al is evaporated with a deposition rate of $2 \, {\rm \AA{}/sec}$. Secondly, a few nm thick Alox layer is created at a constant ${\rm O_{2}}$ pressure of $1.3$~mbar in $11$~minutes. The chamber is then pumped back down to a pressure of $2.2 \cdot 10^{-7}$~mbar. Here, the $75$~nm thick second Al layer is deposited at a rate of $5 \, {\rm \AA{}/sec}$. Lastly, another $11$~minute oxidation step is performed to grow a cap layer over the junctions, as the thin layer is self-terminating. This prevents the junctions from reacting with air when the vacuum chamber is vented.\\
\\
\textbf{Step 9: Aluminum lift-off.} We perform Al lift-off to fabricate the junctions. The chip is placed in N-methyl-2-pyrrolidone (NMP) at $80 \, {\rm ^{\circ}C}$ and left overnight. The solvent strips the resist stack to which the Al is adhered. A pipette is used to flow the hot NMP over the surface, such that the resist and Al film are released from the substrate. The chip is then rinsed twice with IPA. Fig.~\ref{Figure_SI_junction} shows scanning electron microscopy (SEM) images of a Josephson junction made with this recipe. It can be seen that the over-etched substrate is smooth and the step edge onto the transmon island is intact. The designed qubit junctions in this work have a width of $157$~nm and $191$~nm, respectively, from which we expect a SQUID inductance of $4.77$~nH. The average inductance probed of exact copies of the qubit on the same chip is $L_{\rm SQUID} = 4.91$~nH. Therefore, we expect the evaporated junctions to have widths only ${\sim}3\%$ smaller than designed. The size of the SQUID loop is $125 \, {\rm \upmu m^{2}}$.\\
\\
After the fabrication of the microwave chip is finished, it is wirebonded to a printed circuit board (PCB) and ready for the flip-chip assembly. 

\subsection{Membrane Chip}
\textbf{Step 10: Membrane chip with etched phononic shield.} For the top chip of the device, we make use of $200 \, {\rm \upmu m}$ thick Si chips purchased with pre-made phononic shields by our design from Norcada \cite{Norcada}. The ${\rm Si_{3}N_{4}}$ membrane is $300 \, {\rm \upmu m} \times  300 \, {\rm \upmu m}$ wide and $100 \, {\rm nm}$ thick, and is etched to create side walls with an angle of $54.7{\rm ^{\circ}}$. The phononic shield is designed to have a bandgap of $1.4$~MHz wide, centered around $1$~MHz.\\
\\
\textbf{Step 11: Aluminum evaporation.} To couple the membrane capacitively to the microwave cavity, we metallize the membrane with a layer of Al. We perform shadow mask evaporation using a laser-cut Kapton hard mask with a $250 \times 250 \, {\rm \upmu m}$ square hole. A similar evaporation process recipe was used as for the junction deposition. Here, we evaporate once $50$~nm of Al from a direction perpendicular to the chip, and perform an $11$~minute oxidation step.

\subsection{Flip-Chip Device} 
\textbf{Step 12: Device assembly.} In the final fabrication stage, we combine the two chips into a single device using a flip-chip technique. The procedure is performed underneath a microscope, where the membrane chip is flipped upside down on top of the spacers on the bottom chip. The metallized membrane is aligned with the cavity underneath, and the chips are hard-clamped using spring-loaded screws. Further details on the flip-chip holder and device assembly can be found in Ref.~\cite{Soest2025}. The microwave cavity and the qubit are shown before and after the flipping of the membrane in Fig.~\ref{Figure_SI_fridge}(a)~and~(b), respectively.\\
\\
\section{Measurement Setup} \label{SI_Setup}
\begin{figure*}[t!]
    \centering
    \includegraphics[width=0.6\textwidth]{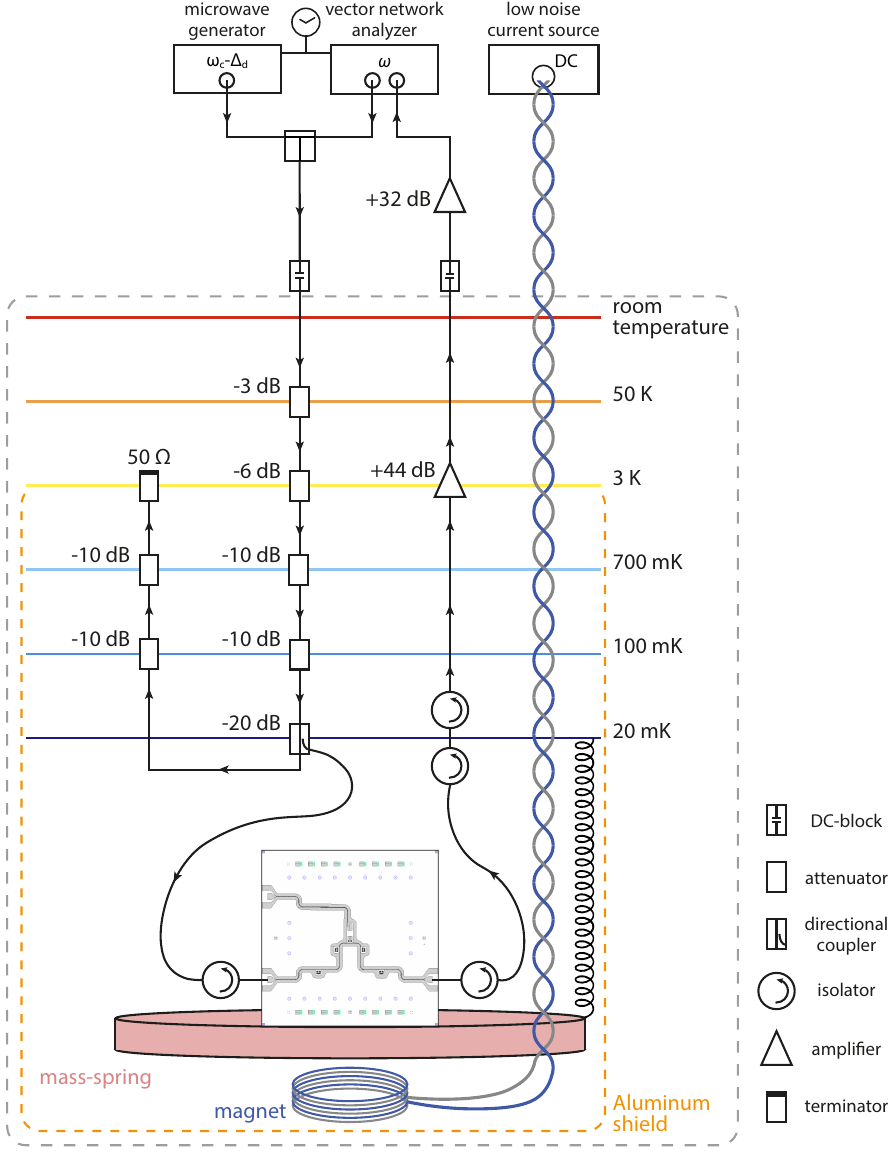}
    \caption{\textbf{Schematic measurement setup.} Detailed information is provided in the text.}
    \label{Figure_SI_wiring}
\end{figure*}
\begin{figure*}[t!]
    \centering
    \includegraphics[width=\textwidth]{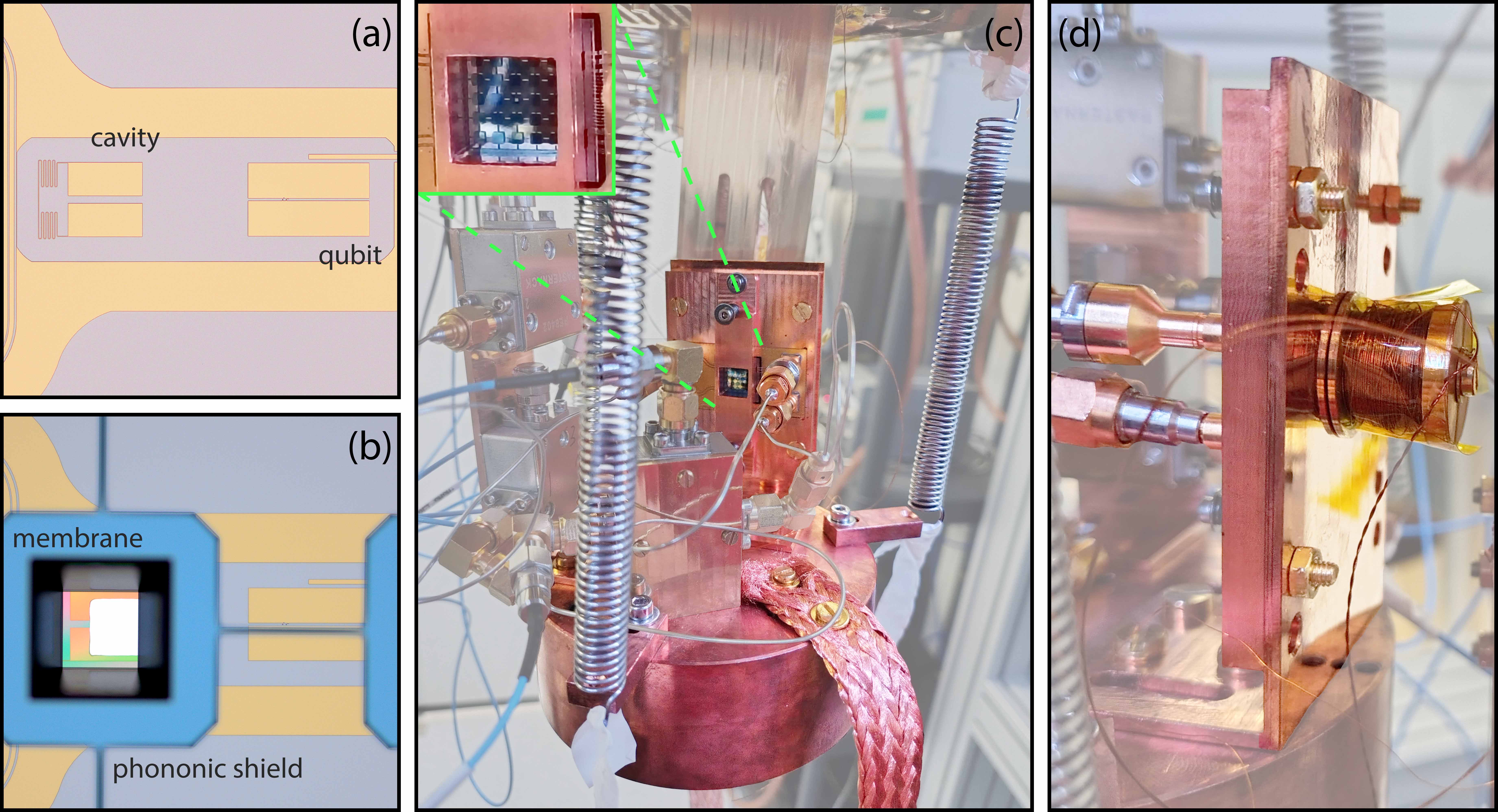}
    \caption{\textbf{Flip-chip device and cryogenic setup for the experiments.} (a) Micrograph of part of the microwave circuit on the bottom chip. NbTiN is yellow, the Si substrate is gray. (b) Micrograph of the same circuit after flipping of the membrane. The Al metallization (white) is aligned with the cavity. Through the ${\rm Si_{3}N_{4}}$ membrane, white light interference can be seen. The phononic shield in the top chip is blue. (c) The flip-chip device is mounted vertically on the mass-spring system and connected with flexible coaxial cables, together with the isolators, temperature sensor and thermal braid. The inset shows the membrane (white square) in the middle of the phononic shield (blue) flipped on top of the microwave chip. (d) The external superconducting magnetic coil mounted at the back of the flip-chip holder.}
    \label{Figure_SI_fridge}
\end{figure*}

%
The experiments in this work were performed in a dilution refrigerator at a base temperature of $T_{\rm base} \approx 20$~mK. The flip-chip device is mounted on a mass-spring system hanging below the mixing chamber plate. A schematic diagram of the experimental setup is shown in Fig.~\ref{Figure_SI_wiring}. The copper mass is thermalized to the base plate through a flexible copper braid. On the mass, a temperature sensor is installed, which is monitored with an AC resistance bridge. Coaxial microwave lines connect the device to the outside of the cryostat. All lines contain DC blocks at room temperature to isolate the system from DC currents and to prevent ground loops. The input line to the device is strongly attenuated. A total of $49$~dB of attenuation was used, on top of the attenuation from the coaxial cables themselves, to block thermal radiation of the line from reaching the device. The final attenuation stage is created by a directional coupler, where most power is led back up, through identical attenuators, to be dissipated at the $3$~K stage. This power dump was installed to enable large pump powers without heating the base plate. Two flexible microwave cables are guided to two isolators, thermalized to and mounted on the mass as close to the flip-chip device as possible. These isolators are installed to reduce the cable resonances we observe over the entire bandwidth. A high-electron-mobility transistor (HEMT) amplifier with a gain of $44$~dB and bandwidth of $4\!-\!8$~GHz was placed on the output line at the $3$~K stage. Two additional isolators were installed between the HEMT and the copper mass to block the thermal radiation of the amplifier reaching the device. Additionally, the qubit on the device is coupled to a coplanar waveguide readout resonator, which was not used in this work. The lines connected to this resonator were $50 \, \Omega\!{\rm -terminated}$, and are omitted from the diagram for clarity. Furthermore, the chip design is shown on the mass-spring system in Fig.~\ref{Figure_SI_wiring}.\\
\\
An external magnetic coil is mounted on the back side of the flip-chip holder, at a distance of $13$ mm from the chip. The coil contains $600$ turns of superconducting wire and is aligned to flux-bias the SQUID loop of the qubit. A DC current can be sent with a low-noise current source, connected with a twisted wire pair. An aluminum shield is installed inside the vacuum can to shield the entire volume below the $3$~K stage from magnetic flux noise. The device and microwave components mounted on the mass are shown in Fig.~\ref{Figure_SI_fridge}(c). The magnet is shown in (d).\\
\\
Outside the refrigerator, we use a single measurement setup for the experiments presented. A vector network analyzer (VNA) is used to send the probe signal to the device. Before reaching the cryostat, the signal passes through a power combiner along with the pump tone from a microwave generator. The signal in the output line of the cryostat first goes through a room-temperature amplifier before returning to the VNA. For all experiments, the employed instruments used a single reference clock.

\section{The Microwave Resonance} \label{SI_Resonances}
\begin{figure}[t!]
    \centering
    \includegraphics{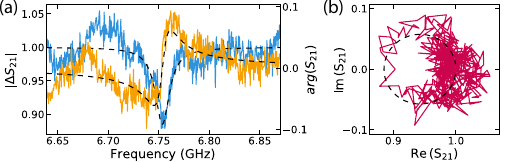}
    \caption{\textbf{Full cavity response.} (a) Single-tone spectroscopy measurement of the flip-chip cavity. The magnitude (phase) response is shown in blue (orange), after applying background correction and phase rotation. (b) The response is shown in the complex plane in red. The Lorentzian fit is shown in both plots as black dashed lines.}
    \label{Figure_SI_cav}
\end{figure}

In the following section, we discuss the fitting routine and background correction of the measured resonances. An \textit{LC}-resonator side-coupled to a transmission feedline can be described by the ideal response function
\begin{equation} \label{EqCoolSI2}
    S_{21}^{\rm ideal}(\omega) = 1 - \frac{\kappa_{\rm ext}}{\kappa_{\rm tot} + 2i\Delta}
\end{equation}
where $\Delta = \omega - \omega_{\rm c}$. No experimental setup, however, is ideal. Therefore, we consider a frequency-dependent complex background to the resonance, caused by amplitude and phase modulation due to the microwave components in out measurement setup. Moreover, we take into account a modification of the cavity response to a Fano-like resonance \cite{Khalil2012, Rieger2023}. The modified response function is given by
\begin{equation} \label{EqCoolSI2}
    \begin{aligned}
        S_{21}^{\rm corrected}(\omega) = 1 + \biggl[ &S_{21}^{\rm ideal}(\omega) \cdot \Bigl( \left( a_{\rm b} + b_{\rm b}\omega + c_{\rm b}\omega^{2} \right) \cdot\\[0.8ex]
        &e^{i\left( d_{\rm b} + e_{\rm b}\omega + f_{\rm b}\omega^{2} \right)} \Bigr) ^{-1} - 1 \biggr] \cdot e^{-i\theta}
    \end{aligned}
\end{equation}%
where we consider an exponential polynomial background function with coefficients $a_{\rm b}\! - \! f_{\rm b}$. Additionally, a rotation in the complex plane is applied by the phase factor $e^{-i\theta}$ with a phase $\theta$.\\
\\
As described in the main text, the flip-chip cavity undergoes such large cavity noise that, without Kerr locking, the frequency fluctuations make the time-averaged cavity very broad. Because the cavity is almost washed out into the background, we cannot perform a reliable fit. However, by performing fast VNA measurements at a timescale shorter than the frequency fluctuations, we are able to obtain the cavity's unperturbed linewidth. The VNA traces were recorded with an IFBW of $10$~kHz and $501$ points, resulting in a measurement time of $213$~ms per trace. Fitting the $500$ traces of Fig.~3(a) individually, we acquire a linewidth of $\kappa_{\rm tot} / 2\pi \approx 15$~MHz. The corrected response of one of these traces, together with a Lorentzian fit is shown in Fig.~\ref{Figure_SI_cav}.

\section{Interchip Distance Estimation} \label{SI_Simulation}
\begin{figure}[t!]
    \centering
    \includegraphics{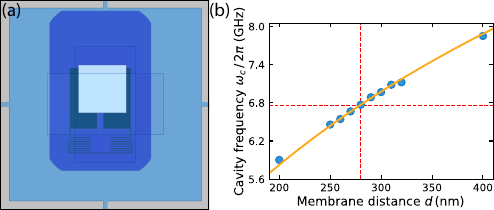}
    \caption{\textbf{Simulated cavity resonance frequency for different gap sizes.} (a) Top view of the geometry used in the \textit{Ansys HFSS} eigenmode simulation, matching the exact misalignment of the experimental device. The superconducting cavity is shown in black, the groundplane in gray. The superconducting metal on the membrane is visible as the white square, and the silicon phononic shield as the light blue structure. (b) Simulated eigenfrequency of the microwave cavity for a parametric sweep of the distance $d$ between the two chips. Simulated frequencies are shown as blue circles, and the fit as the orange line. The red dashed lines correspond to the frequency measured in the device, and the corresponding gap size of $d = 280$~nm.}
    \label{Figure_SI_sim}
\end{figure}

After fabrication of the microwave and membrane chips, the flip-chip device was assembled inside a cleanroom. Due to the small misalignment in the device, we expect a slight correction in the cavity frequency. An adjusted configuration was made in the finite element simulation software \textit{Ansys HFFS}. Based on the micrograph shown in Fig.~\ref{Figure_SI_fridge}(b), the exact misalignment of the metal on the membrane, and with respect to the cavity underneath, was incorporated. The geometry of the phononic shield and the etched membrane were included to account for the relative permittivity of the silicon top chip. A top view of the simulation geometry is shown in Fig.~\ref{Figure_SI_sim}(a). The distance between the two chips $d$ was parametrically swept while performing eigenmode simulations. This gives us a corrected model to compare with the measured distance-dependent resonance frequency, as shown in Fig.~\ref{Figure_SI_sim}(b). The misalignments result in a slightly higher cavity frequency for the same distance, as compared to the case of perfect alignment. The simulation data is fit with $\omega_{\rm c} = 1 / \sqrt{LC_{\rm tot}}$, where $C_{tot}$ includes the misalignment as given by
\begin{equation} \label{Eq_freq_parallel_plate}
    C_{\rm tot} = C_{\rm pads} + \frac{\epsilon_{0}}{d} \cdot \frac{A_{1}A_{2}}{A_{1} + A_{2}}
\end{equation}
where $\epsilon_{0}$ is the vacuum permittivity, and $A_{1,2}$ are the overlapping areas of each pad with the membrane. $C_{\rm pads}$ is the loaded capacitance between the two pads of the cavity. The average measured frequency of the cavity is $\left< \omega_{\rm c} \right> / 2\pi = 6.76$~GHz, from which we estimate a distance between the chips of $d = 280$~nm, as indicated by the red dashed lines. This gap size is in agreement with the model for white light interference in Ref.~\cite{LingeJohnsen2018}.

\section{Dispersive Cavity-Qubit Coupling} \label{cQED}
\begin{figure*}[t!]
    \centering
    \includegraphics{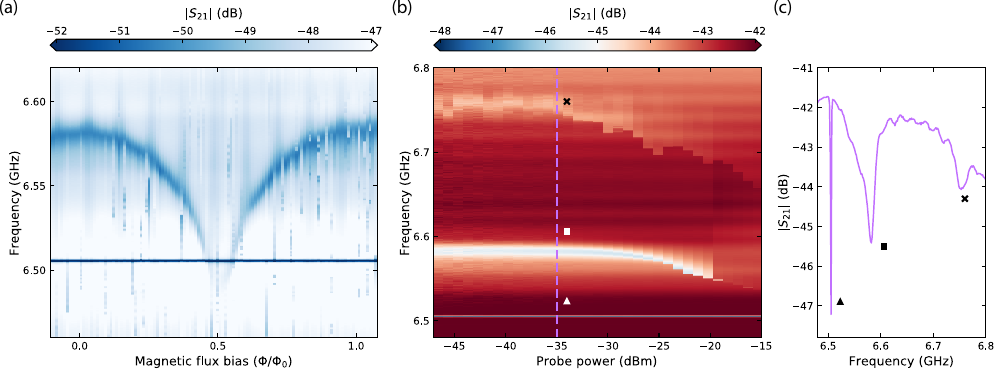}
    \caption{\textbf{Nonlinear resonator response due to dispersive coupling to a transmon qubit.} (a) Magnitude of the single-tone transmission response $|S_{21}|$ as a function of the applied magnetic flux $\Phi$, obtained by sweeping the current through the external magnetic coil. The reference resonator at $\omega_{\rm ref}/2\pi = 6.505$~GHz is insensitive to the applied flux, whereas the qubit readout resonator undergoes a periodic frequency shift due to its dispersive coupling to the transmon qubit. The flip-chip cavity exhibits a rapidly varying periodic frequency shift and is operated at the maximum of the global flux arch for the experiments presented in the main text. (b) Magnitude of $|S_{21}|$ as a function of probe power. The reference resonator response remains linear, while the readout resonator exhibits a Kerr-induced AC Stark shift and Duffing-type nonlinear response. The flip-chip cavity frequency likewise shifts downward with increasing probe power, develops an asymmetric lineshape, and enters the bistable regime at high powers. (c) Single-tone transmission spectrum at a probe power of $-35$~dBm, corresponding to the dashed line in panel (b), identifying the reference resonator (\scalebox{1.2}{$\blacktriangle$}), the qubit readout resonator ($\blacksquare$), and the flip-chip cavity (\ding{54}).}
    \label{Figure_SI_sweeps}
\end{figure*}

The microwave circuit on the bottom chip of the device contains several additional resonators beyond the cavity studied in the main text. The qubit is capacitively coupled to the cavity and to a readout resonator. In addition, several ``reference" resonators are coupled to the transmission line. All measurements presented in this work are performed through this transmission line. In Fig.~\ref{Figure_SI_sweeps}(a) shows single-tone spectroscopy of the system as a function of the applied external magnetic flux, obtained by sweeping the current in the external magnetic coil. Increasing the nonlinear inductance of the Josephson junctions reduces the qubit frequency. This reduces the detuning between the qubit and the coupled resonators, thereby enhancing their effective self-Kerr nonlinearities. Three resonances are observed within the shown frequency range. One of the reference resonators, at $\omega_{\rm ref}/2\pi = 6.505$~GHz, appears as a narrow transmission dip and exhibits no dependence on the applied magnetic flux. The readout resonator at $\omega_{\rm ro}\left( \Phi\!=\!0 \right) /2\pi = 6.58$~GHz displays a significant periodic frequency shift \cite{Koch2007, Blais2021}. The cavity subject to frequency noise shows a rapidly varying periodic frequency shift of the cavity, likely caused by a flux-dependent force \cite{Song2009, Kremen2016, Sahu2022, Luschmann2022}. For the experiments presented in this article, the cavity is tuned to its maximum frequency at the sweet spot of the global flux arch to minimize the frequency fluctuations.

Figure~\ref{Figure_SI_sweeps}(b) shows the single-tone response of the same three resonances as the probe power at the room-temperature end of the measurement chain is increased. The spectrum at a probe power of $-35$~dBm is shown in Fig.~\ref{Figure_SI_sweeps}(c), where the three resonances are identified. The reference resonator remains unchanged, as its circuit elements are linear with respect to the intraresonator photon number. In contrast, the qubit readout resonator exhibits a downward frequency shift with increasing probe power, as it inherits a self-Kerr nonlinearity through its dispersive coupling to the qubit. The magnitude of this shift depends on the photon number $n$ in the resonator. At sufficiently large $n$, the resonator enters a regime of strongly asymmetric response, consistent with the equations of motion of a Duffing resonator. The third resonance in this frequency range is identified as the flip-chip cavity, centered around $\omega_{\rm c} = 2\pi \cdot 6.76$~GHz and exhibiting significant frequency fluctuations. The observed time-averaged linewidth substantially is increased. Its self-Kerr nonlinearity likewise produces a downward frequency shift with increasing probe power, accompanied by an increasingly asymmetric resonance lineshape. Upon entering the bistable regime, the cavity response splits into low- and high-amplitude branches, corresponding to low and high intracavity photon occupation.

\section{Fitting Kerr from Signal Frequency}
\begin{figure}[t!]
    \centering\includegraphics{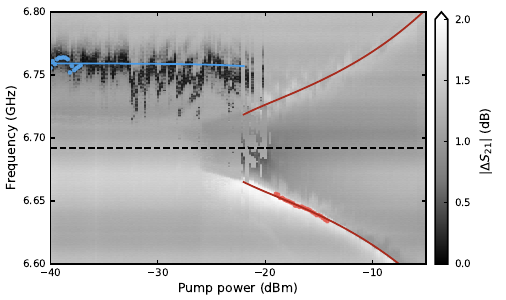}
    \caption{\textbf{Inferring the cavity Kerr by fitting the signal frequency shift} The $S_{21}$ dataset from Fig.~2 is shown alongside the theoretically predicted signal and idler frequency shifts for a cavity Kerr of $ K_{\rm c}/2\pi = -24$~kHz. This is consistent with a total line attenuation of ${\sim}58.7$~dB. The dashed horizontal line represents the pump frequency, which is placed at $\Delta_{\rm p}/2\pi = -67.3$~MHz below the bare cavity resonance frequency. The red (blue) markers indicate the fit resonance frequencies of the locked signal mode (unlocked cavity mode) frequency. The solid lines are theoretical curves based on Eq.~\ref{Eqn:F3}, representing the signal and idler frequencies at which the cavity photon number is in the low-photon branch (blue) and the high-photon branch (red).} \label{Fig:S17}
\end{figure}

To determine the Kerr nonlinearity of the pumped cavity, we measured the small-signal response of the device as a function of probe frequency and applied pump power. The pump was applied at a fixed frequency $\omega_{\rm p}/2\pi = 6.692$ GHz while the probe frequency was swept through the signal/idler response of the pumped Kerr mode, see Fig.~\ref{Fig:S17}.

The pumped cavity was modeled as a single Kerr oscillator with resonance frequency $\omega_{\rm c }$, total decay rate $\kappa$, external coupling rate $\kappa_{\rm ext}$, and Kerr coefficient $\mathcal{K}_{\rm c}$. For a coherent pump field with incident photon flux $\vert S \vert^2$, the steady-state intracavity photon number $n_{\rm c}$ satisfies
\begin{equation}
    n_{\rm c}\bigg[ (\Delta_{\rm p} - \mathcal{K}_{\rm c}n_{c})^2 + (\kappa/2)^2  \bigg] = (\kappa_{\rm ext}/2)\vert S \vert^2,
\end{equation}
where $\Delta_{\rm p} = \omega_{\rm p} - \omega_{\rm c}$ is the pump detuning from the bare cavity resonance. The steady-state photon number is obtained by solving the cubic equation
\begin{equation}
\begin{split}
    \mathcal{K}_{\rm c}^2 n_{\rm c}^3 &- 2\mathcal{K}_{\rm c}\Delta_{\rm p} n_{\rm c}^2 + \bigg[ \Delta_{\rm p} + (\kappa/2)^2\bigg]n_{\rm c}\\ &-(\kappa_{\rm ext}/2)\vert S \vert^2 =0.
\end{split}
\end{equation}
For each applied pump power, the positive real roots of this cubic equation were calculated. When multiple physical roots were present, the low- and high-photon-number solutions were retained and labeled $n_{\rm LB}$ and $n_{\rm HB}$, respectively. The intermediate root is an unstable solution and is ignored.

At low pump powers, the cavity mode is initially attracted to the pump, while the system remains in the low-photon branch. However, abruptly upon crossing an exceptional point, the weak linear probe measures a pair of signal-idler modes symmetric about the strong pump tone \cite{Sani2021}. These modes have frequencies
\begin{equation}
    \omega_{\pm}(n_{\rm c}) = \omega_{\rm p} \pm \sqrt{(\Delta_{\rm p} - \mathcal{K}_{\rm c}n_{c})(\Delta_{\rm p}- 3\mathcal{K}_{\rm c}n_{c})}.
    \label{Eqn:F3}
\end{equation}
The two signs correspond to the two conjugate modes of the pumped Kerr resonator, separated symmetrically about the pump in the rotating-frame description. In the fit, the locked branch was modeled using the high-amplitude steady-state solution $n_{\rm HB}$.

At low pump powers, before the locked branch appears, the resonance frequency was extracted from a Lorentzian fit of the background-subtracted magnitude trace. The bare cavity resonance used in the model was determined as the mean of the first 15 low-power-extracted resonance frequencies, yielding $\omega_{\rm c} / 2\pi = 6.759$~GHz. The Kerr coefficient $\mathcal{K}_{\rm c}$ was obtained by fitting the measured locked-branch frequencies to Eq.~\ref{Eqn:F3}. Where $n_{\rm c} = n_{\rm HB}$ is the high-amplitude steady-state photon amplitude of the Kerr oscillator. The fit produced a Kerr coefficient of $\mathcal{K}_{\rm c}/2\pi = -24.1$~kHz, see Fig.~\ref{Fig:S17}.

\section{Numerical Simulations of the Locking Mechanism}
To model the observed suppression of cavity-frequency fluctuations, we simulate a driven Kerr cavity in which the bare resonance frequency is weakly modulated over time. In the frame of the pump, the intracavity field $\alpha(t)$ obeys
\begin{equation}
    \dot{\alpha}(t) = \bigg[ i (\Delta_{\rm p}(t) - \mathcal{K}_{\rm c}\vert \alpha(t)\vert^2) - \kappa/2 \bigg]\alpha(t) + S.
    \label{Eqn:G1}
\end{equation}
Where $\kappa$ is the total linewidth of the resonator, $\mathcal{K}_{\rm c}$ is the Kerr coefficient, $S$ is the pump amplitude, and $\Delta_{\rm p}(t) = \omega_{\rm p} - \omega_0(t)$ is the instantaneous detuning between the pump frequency and the bare cavity frequency. 

To probe the frequency dependence of the locking mechanism, we impose a small sinusoidal modulation of the bare cavity frequency,
\begin{equation}
    \omega_0(t) = \bar{\omega}_0 + \delta \omega_0\cos(\omega_{\rm m}t),
\end{equation}
where $\omega_{\rm m}$ is the modulation frequency, and $\delta\omega_0$ is sufficiently small such that the system remains in the same dynamical branch. 
\begin{figure*}[t!]
    \centering\includegraphics[width=\linewidth]{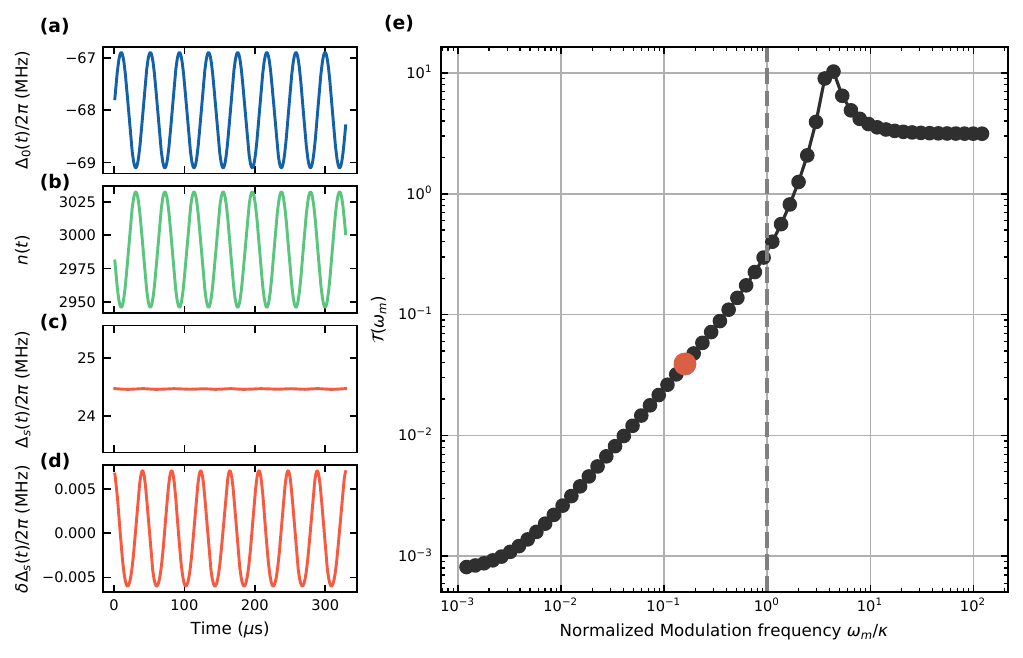}
    \caption{\textbf{Kerr-cavity simulations show that slow cavity fluctuations are suppressed in the signal mode.} (a) Sinusoidal modulation of bare cavity resonance $\Delta_{\rm 0}(t) = \omega_{\rm p} - \omega_0(t)$. (b) Resulting modulation of the intracavity photon number, $n(t)$, due to the nonlinear Kerr response. (c) Induced signal-mode detuning, $\Delta_{\rm s}(t) = \omega_{\rm p} - \omega_{\rm s}(t)$, shown on the same scale as (a). (d) Zoomed-in view of the signal-mode fluctuations $\delta\Delta_{\rm s}(t)$. (e) Extracted transfer function $\mathcal{T}(\omega_{\rm s})$ versus modulation frequency. The dashed vertical line marks $\omega_{\rm m} = \kappa$, and the red point represents the data shown in (a-d). For modulation frequencies below the cavity linewidth, the photon number adiabatically follows the pump-detuning, and therefore the signal-frequency fluctuations are strongly suppressed.}
    \label{Figure_SI_numerics}
\end{figure*}
For each modulation frequency, we numerically integrate Eq.~\ref{Eqn:G1} and calculate the instantaneous intracavity photon number $n(t) = \vert \alpha(t)\vert^2$. From this, we compute the instantaneous signal/idler mode frequencies using Eq.~\ref{Eqn:F3}. We then extract the amplitude $\delta\omega_{\rm s}(t)$ of the oscillation of the signal-mode frequency at the modulation frequency $\delta\omega_0(t)$. This defines a frequency-noise transfer function
\begin{equation}
    \mathcal{T}(\omega_{\rm m}) = \bigg\vert \frac{\delta\omega_{\rm s}(\omega_{\rm m})}{\delta\omega_0} \bigg\vert.
\end{equation}

A value $\mathcal{T}<1$ indicates suppression of the bare cavity-frequency noise in the dressed signal mode. In the low-frequency limit, the intracavity field adiabatically follows the modulation, allowing the Kerr-induced change in photon number to partially compensate the bare cavity-frequency shift. This produces an effective negative feedback and suppresses the observed frequency fluctuations. For modulation frequencies approaching and exceeding the cavity linewidth, $\omega_{\rm m} > \kappa$, the intracavity field can no longer respond adiabatically, and the suppression weakens. Fig.~\ref{Figure_SI_numerics} shows the simulated transfer function $\mathcal{T}(\omega_{\rm s})$, demonstrating that noise is strongly reduced for modulation frequencies below the cavity linewidth and that the suppression rolls off at frequencies of order $\kappa$. Fig.~\ref{Figure_SI_numerics}(c,d) shows the residual error in the locked dressed cavity resonance for the modulation frequency corresponding to the red point in (e). A small residual error is expected for purely proportional feedback with finite gain. Perfect cancellation of static or low-frequency fluctuations would require an integral feedback component.

\section{Cavity Noise Model and Allan Deviation}
\begin{figure*}[t!]
    \centering
    \includegraphics{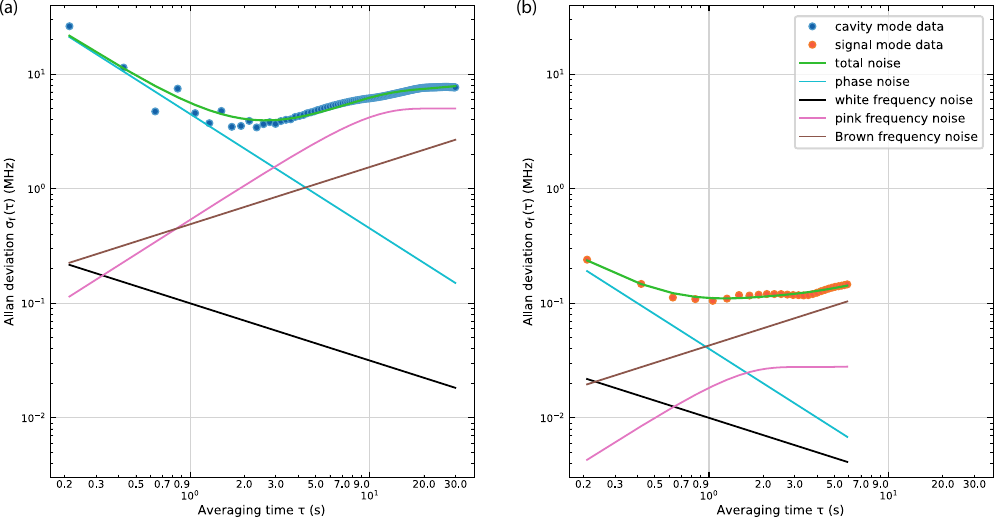}
    \caption{\textbf{Fit noise model of the Allan deviations.} The unlocked (blue circles in (a)) and locked (red circles in (b)) overlapping fractional frequency Allan deviation. The total fit model is shown in green. The individual noise contributions are shown in cyan (phase noise), black (white frequency noise), pink ($1/f$ frequency noise), and brown (Brownian frequency noise). With the feedback on, each contribution is reduced by one to two orders of magnitude.}
    \label{Figure_SI_Allan_deviation}
\end{figure*}%
The Allan variance is a statistical tool originally developed to analyze the stability of atomic clocks. It can be used to measure the fractional frequency fluctuations of a harmonic oscillator undergoing noise in its resonance frequency. The Allan variance for a fractional resonance frequency $x = \left( f - \left< f \right> \right) / \left< f \right>$ is given by

\begin{equation}
    \sigma_{\rm f}^{2}\left( \tau \right) = \frac{1}{2\left( N - 1 \right)} \sum_{i=1}^{N-1} \left( x_{i+1} - x_{i} \right)^{2}
\end{equation}
where $N$ is the measured number of resonance frequencies of the harmonic oscillator \cite{Barnes1971, Riley2008}. The observation time $\tau$ is the time in which consecutive data points are averaged. $\tau$ can thus be swept from the sampling time of the measurement $t_{\rm sampling}$ to half of the total measurement time. Taking the square root of the Allan variance gives the Allan deviation $\sigma_{f}$. Where a simple standard deviation is sufficient when the noise source is purely white, the Allan deviation is the superior analysis method when the system suffers from a variety of noise components.\\
\\
However, as the number of averaged data points becomes small for increasing $\tau$, the uncertainty of the normal Allan deviation increases significantly. Therefore, one can instead perform the overlapping Allan deviation, where all possible combinations of the data points are used. This results in larger time-averaged datasets, and thereby lower uncertainty. The overlapping Allan deviation is given by

\begin{equation}
    \sigma_{\rm f}\left( \tau \right) = \sqrt{\frac{1}{2m^{2}\left( N - 2m + 1 \right)} \sum_{j=1}^{N - 2m + 1} \left[ \sum_{i=j}^{j + m - 1} \left( x_{i+m} - x_{i} \right)^{2} \right] }
\end{equation}
where $m$ is the averaging factor such that $\tau = m \cdot t_{\rm sampling}$ \cite{Howe1981}. The overlapping Allan deviation is computed for the fractional frequencies found in the fits of the resonances. We then renormalize the found deviation by converting back to absolute frequency units. This is shown as $\sigma_{\rm f} \left( \tau \right)$ in Fig.~\ref{Figure_SI_Allan_deviation} for the unlocked cavity (blue circles in (a)) and the Kerr-locked signal mode (red circles in (b)). 

The observed instantaneous frequency $f_{\rm inst}(t)$ is determined by both temporal fluctuations of the cavity $f_{\rm c}(t) = \omega_{\rm c}(t) / 2\pi$ together with the fluctuations of the phase $\phi(t)$  of the output signal using the relation:
\begin{equation}
    f_{\rm inst}(t) = f_{\rm c}(t) + \frac{1}{2\pi}\frac{d\phi(t)}{dt}
\end{equation}
To analyze the found Allan deviations, we fit this to our noise model, consisting of multiple types of noise \cite{Rubiola2005, Riley2008}. We consider a combination of white and $1/f$ (pink) phase noise, as well as white, pink and Brownian frequency noise. The power spectral density of this sum can be written as
\begin{equation}
    \begin{aligned}
        S_{\rm tot}(f) &= \left(2\pi f \right)^{2} C_{\rm white\,phase}^{2} + \left(2\pi f \right) C_{\rm pink\,phase}^{2} \\[0.8ex]
        &+ C_{\rm white}^{2} + \frac{C_{\rm pink}^{2}}{2\pi f} + \frac{C_{\rm Brown}^{2}}{\left(2\pi f\right)^{2}}
    \end{aligned}
\end{equation}
where $C_{\eta}$ are the coefficients of each noise contribution $\eta$. We can calculate the variance that corresponds to each power spectral density of noise as

\begin{equation}
    \sigma_{\rm \eta}^{2} \left( \tau \right) = \int_{0}^{\infty} S_{\rm \eta}\left( f \right) \left| H \left( f \right) \right|^{2} \,df
\end{equation}
where $\left| H \left( f \right) \right|^{2}$ is the transfer function related to the specific variance being calculated. For the Allan deviation, this equation becomes

\begin{equation}
    \sigma_{\rm \eta}^{2} \left( \tau \right) = 2\int_{0}^{\infty} S_{\rm \eta}\left( f \right) \frac{\sin^{4} \left( \pi f \tau \right)}{\left( \pi f \tau \right)^{2}} \,df
    \label{Eq_sigma_transfer}
\end{equation}

In the following, we consider each noise contribution in turn, starting with white frequency noise. As the power spectral density $S_{\rm white}\left( f \right) = C_{\rm white}^{2}$ is independent of frequency, the Allan deviation is given by

\begin{equation}
    \sigma_{\rm white} \left( \tau \right) = \frac{C_{\rm white}}{\sqrt{\tau}}
\end{equation}

This results in a slope $-\frac{1}{2}$ on a typical log-log Allan deviation plot. Next, we will consider the Brownian noise term. Solving Eq.~\ref{Eq_sigma_transfer}, this gives

\begin{equation}
    \sigma_{\rm Brown} \left( \tau \right) = \sqrt{\frac{\tau}{3}}C_{\rm Brown}
\end{equation}

This noise term is characterized by a slope of $\frac{1}{2}$. The $1/f$ noise contribution is more subtle to model. As for many physical systems, flicker noise becomes negligible above a certain corner frequency $f_{0}$. We can therefore define the power spectral density as \cite{IEEE1998}

\begin{equation}
    S_{\rm pink}\left( f \right) =
    \begin{cases}
        \displaystyle \dfrac{C_{\rm pink}^{2}}{2\pi f} & f \le f_{0} \\
        \displaystyle 0 & f > f_{0}
    \end{cases}
\end{equation}

Inserting this power spectral density in the integral to transform to the Allan deviation gives

\begin{equation}
    \sigma_{\rm pink}(\tau)  = \sqrt{
    \begin{aligned}
        \frac{2 C_{\rm pink}^{2}}{\pi} \cdot \biggl[
        & \ln(2) - \frac{\sin^{3}(\pi f_{0} \tau)}{2 (\pi f_{0} \tau)^{2}} \\
        & \cdot \Bigl( \sin(\pi f_{0} \tau) + 4 (\pi f_{0} \tau)\cos(\pi f_{0} \tau) \Bigr) \\
        & + \operatorname{Ci}(2\pi f_{0} \tau) - \operatorname{Ci}(4\pi f_{0} \tau) \biggr]
    \end{aligned}
    }
\end{equation}

where $\operatorname{Ci} \left( x \right)$ is the cosine integral function

\begin{equation}
    \operatorname{Ci} = \int_{0}^{x} \frac{1 - \cos\left( u \right)}{u} \,du
\end{equation}

For short averaging times $\tau$, this Allan deviation scales linearly with $\tau$. At longer averaging times, this levels off to a value of $\sqrt{\frac{2\ln(2)}{\pi}} \, C_{\rm pink} \approx 0.664 \, C_{\rm pink}$, resulting in a flat region in the log-log representation.\\
\\
Finally, we will treat the derivation of phase noise. We will consider white and pink phase noise contributions separately. Inserting the power spectral density of white phase noise into Eq.~\ref{Eq_sigma_transfer} results in

\begin{equation}
    \sigma_{\rm white\,phase}^{2} = \frac{16 C_{\rm white\,phase}^{2}}{\tau^{2}} \int_{0}^{\infty} \sin^{4}\left( \pi f \tau \right) \,df
\end{equation}

This integral diverges for large frequencies and therefore cannot represent a physical system. In order to solve for this problem, we define a measurement bandwidth $f_{1}$ and compute the integral as

\begin{equation}
    \sigma_{\rm white\,phase} = \frac{\sqrt{3f_{1}}}{\pi\tau}C_{\rm white\,phase}
\end{equation}

On the log-log Allan deviation plot, this results in a slope of $-1$. However, we have to be careful when comparing this with pink phase modulation of the cavity resonance, which is derived as follows. 

\begin{equation}
    \sigma_{\rm pink\,phase}^{2} = \frac{8 C_{\rm pink\,phase}^{2}}{\pi\tau^{2}} \int_{0}^{\infty} \frac{\sin^{4}\left( \pi f \tau \right)}{f} \,df
\end{equation}

In order to solve the problem of the diverging integral, we again define the measurement bandwidth $f_{1}$ and obtain

\begin{equation}
    \sigma_{\rm pink\,phase} = \frac{\sqrt{3}}{\sqrt{\pi}\tau} \ln\left( \pi f_{1} \tau \right) \cdot C_{\rm pink\,phase} + \cdots
\end{equation}

where we neglect a constant and higher-order factors. In the log-log representation of the Allan deviation, the logarithmic term is a small deviation from the $-1$ slope. In realistic experiments, this will be indistinguishable. It is therefore common to treat white and pink phase noise as a single contribution when interpreting an overlapping fractional frequency Allan deviation.\\
\\
In Fig.~3 we performed the Allan deviation $\sigma_{\rm f} \left( \tau \right)$ on the fractional resonance frequencies found with the fits of the fast VNA traces, of both the cavity and the signal mode. The shown Allan deviation is converted back to absolute frequency units. In Fig.~\ref{Figure_SI_Allan_deviation} these results are shown together with the individual noise components extracted from our fit model. For small $\tau$, the unlocked bare cavity in (a) is dominated by phase noise. As $\tau$ increases, Brownian noise becomes dominant. This results in the total Allan deviation increasing immediately after the $-1$ slope, before reaching the $1/f$ noise floor.\\
\\
The fit of the Allan deviation of the locked signal mode in Fig.~\ref{Figure_SI_Allan_deviation}(b) shows that all noise components are reduced, compared to the unlocked bare cavity. With the Kerr-feedback, the system does reach the flat flicker noise, which is expected in the presence of instrumental noise in the experiment. Table~\ref{Table_noise_coef} shows the individual noise coefficients $C_{\eta}$ obtained from the fits for the unlocked bare cavity and the locked signal mode. 
The last column shows the suppression of each noise contribution for when the intrinsic feedback mechanism is used, compared to the noisy cavity resonance. The total cavity noise is reduced by almost two orders of magnitude.\\
\\
\begin{table}[ht]
    \centering
    \renewcommand{\arraystretch}{1.4}
    \setlength{\tabcolsep}{10pt}
    \begin{tabularx}{0.5\textwidth}{ABCD}
        \hline
        \rule{0pt}{4ex}%
        \textnormal{\textbf{noise}} &
        \textnormal{\textbf{unlocked} $C_{\eta}$} &
        \textnormal{\textbf{locked} $C_{\eta}$} &
        \textnormal{\textbf{ratio} $\dfrac{\sigma_{\rm f, \, unlocked}}{\sigma_{\rm f, \,  locked}}$} \\
        \midrule
        {\rm white} & 0.100 & 0.010 & 9.92\\
        {\rm pink}  & 7.65  & 0.042 & 179\\
        {\rm Brown} & 0.848 & 0.074 & 25.8\\
        {\rm phase} & 4.50  & 0.040 & 110\\
        {\rm total} &       &       & 52.1\\
        \bottomrule
    \end{tabularx}

\caption{Coefficients of each noise contribution before and after applying Kerr-feedback, extracted from the fits. The rightmost column shows the ratio of between the Allan deviations of the unlocked cavity and the locked signal mode.}
\label{Table_noise_coef}
\end{table}

\end{document}